\documentclass[preprint,showpacs,preprintnumbers,amsmath,amssymb]{revtex4}

\usepackage{graphicx}
\usepackage{dcolumn}
\usepackage{bm}
\usepackage{hyperref}
\hypersetup{colorlinks=false}
\newcommand{\fet} [1]{\mbox{\boldmath $#1$}}

\begin{document}


\title{The three dimensional calculations of $NN$ bound and scattering states with chiral potential up to N$^3$LO}

\author{S. Bayegan}%
 \email{bayegan@khayam.ut.ac.ir}
 \author{M. A. Shalchi}
\email{shalchi@khayam.ut.ac.ir}
\author{M. R. Hadizadeh}
\email{hadizade@khayam.ut.ac.ir}

\affiliation{Department of Physics, University of Tehran, P.O.Box
14395-547, Tehran, Iran
}%

\date{\today}

\begin{abstract}

The recently developed chiral nucleon-nucleon ($NN$) potential by Epelbaum \emph{et al.} has been employed to
study the two-nucleon bound and scattering states. Chiral $NN$ potential up to next-to-next-to-next-to leading order (N$^3$LO) is used to calculated the np differential cross section and deuteron binding energy in a realistic three dimensional approach. The obtained results based on this helicity representation are compared to the standard partial wave (PW) results. This comparison shows that the 3D approach provides the same accuracy in the description of $NN$ observables and the results are in close agreement with available experimental data.

\end{abstract}

\pacs{ 21.45.-v, 21.45.Bc, 13.75.Cs, 21.10.Hw }

\keywords{Suggested keywords}
\maketitle

\section{Introduction} \label{sec:introduction}

It is a long time that the standard PW decomposition has been used to solve the few-body problems.
In this approach one should sum all PW's to infinite order, but in practice
one truncates the sum to a finite angular momentum number which is
dependent to the considered energy. It means that in higher
energies one will need many PW components, which leads to very complicated expressions, to achieve the convergence results.
It appears therefore natural to avoid the very involved angular momentum algebra which is inherent in the PW representation of permutations, transformations and especially the $3N$ forces and work directly with vector variables \cite{Rice-FBS14}.
To this aim in the past
decade the main steps have been taken by Ohio-Bochum collaboration (Elster, Gl\"{o}ckle \emph{et al.})
and Bayegan \emph{et al.} to implement the 3D approach in few-body bound
and scattering calculations (see for examples Refs. \cite{Fachruddin-PRC62}-\cite{Bayegan-NPA814}).
The 3D approach replaces the discrete angular momentum quantum
numbers with continuous angle variables and consequently it
takes into account automatically all PW's. So in contrast to the truncated PW
approach, the number of equations in the non truncated 3D
representation is energy independent. Therefore this non PW method
is more efficient and applicable to the three- and four-nucleon
scattering problems which consider higher energies than the
corresponding bound state problems.
It should be clear that the building blocks
to the few-body calculations without angular momentum
decomposition are two-body off-shell t-matrices, which depend on the magnitudes of the initial and
final Jacobi momenta and the angle between them. Fachruddin \emph{et al.} have formulated the $NN$ bound and scattering states in a 3D representation and they have numerically illustrated the np differential cross section and deuteron binding energy by using two realistic
model interactions, i.e. the Bonn-B and the AV18 \cite{Fachruddin-PRC62,Fachruddin-PRC63}. They have incorporated
the momentum vectors directly into the bound and scattering equations, and the
total spin of the two nucleons is treated in a helicity
representation with respect to the relative momentum of the two nucleons.
Despite many successes that conventional approaches achieved in incorporating the $NN$ potentials, like the CD-Bonn, the Nijmegen I and II, and the AV18, in nuclear structure and reaction calculations, there are certain deficiencies that require a reliable approach which is based on the theory of strong interactions, the quantum chromodynamics (QCD). These deficiencies can be categorized as no connection to QCD, model-dependent with the lack of $3N$ force to be add on, the gauge and chiral symmetries hard to be reached and finally fine tuning in not achievable order by order of increasing momenta. Based on the spontaneously and explicitly broken chiral symmetry it is possible to construct nuclear forces in the framework of the chiral perturbation theory. This approach has been
founded by Weinberg \cite{Weinberg-PRLB251,Weinberg-NPB363} and further expanded
by Ord\'{o}\~{n}ez \emph{et al.} \cite{Ordonez-PRLPR}, Kaiser \emph{et al.}
\cite{Kaiser-NPPR}, Entem \emph{et al.} \cite{Entem-PRC68}, and recently by Epelbaum
\emph{et al.} \cite{Epelbaum-EPJA19, Epelbaum-RMP}.
In order to compare the 3D and the PW approaches in a more
fundamental basis, we intend to incorporate the new chiral potential \cite{Epelbaum-NPA747} into the 3D few-body calculations. In the first step we are preparing this potential in an appropriate operator form, which is consistent with 3D representation, to calculate the np differential cross section and also the deuteron binding energy.

\section{A brief review of the 3D formalism for $NN$ bound and scattering states} \label{sec: NN formalism}

The $NN$ differential cross section is given as:
\begin{eqnarray} \label{eq.sigma}
&& \frac{d\sigma}{d\Omega}= (2\pi)^{4} \, (\frac{m}{2})^{2} \sum_{ m'_{t_{1}}, \, m'_{t_{2}}, \, m_{t_{1}}, \, m_{t_{2}}}
\nonumber \\ && \biggl| \, _{a}\langle {\bf p}' \, m_{s_{1}} m_{s_{2}} \, m'_{t_{1}}
m'_{t_{2}}  |T| {\bf p} \, m_{s_{1}} m_{s_{2}} \, m_{t_{1}} m_{t_{2}} \rangle_{a} \biggr |^{2},
\end{eqnarray}
where $m_{s_{i}}$ and $m_{t_{i}}$ indicate the projection of the spin and isospin of
the nucleons, ${\bf p}$ and ${\bf p}'$ are initial and final relative momentum of the two nucleons and the operator $T$ is the $2N$ transition matrix determined by Lippmann-Schwinger equation.
In order to calculate the $NN$ differential cross section we need to calculate the matrix elements of the physical
representation of $NN$ $T$-matrix, i.e.:
\begin{eqnarray} \label{eq.Ta}
_{a}\langle {\bf p}' \, m_{s_{1}} m_{s_{2}} \, m'_{t_{1}}
m'_{t_{2}} |T| {\bf p} \, m_{s_{1}} m_{s_{2}} \, m_{t_{1}} m_{t_{2}} \rangle _{a},
\end{eqnarray}
which are given in the antisymmetrized basis states, i.e. $| {\bf p} \, m_{s_{1}} m_{s_{2}} \, m_{t_{1}} m_{t_{2}} \rangle _{a} \equiv \frac{1}{\sqrt{2}}(1-P_{12}) | {\bf p} \, m_{s_{1}} m_{s_{2}} \, m_{t_{1}} m_{t_{2}} \rangle.$
These matrix elements can be obtained by a summation over the on-shell
momentum helicity $T$-matrices multiplied with the rotational matrices and Clebsch-Gordon coefficients \cite{Fachruddin-PRC68}.
As indicated in Ref. \cite{Fachruddin-PRC63} the projection of the Schr\"{o}dinger equation on the helicity basis states leads to the coupled integral equations which, after simplification, are actually only two dimensional integral equations.
Since the deuteron has spin $1$ there are three possible values for the helicity projections, namely $ \Lambda =-1,\, 0,\, +1 $. The symmetry properties allow to consider only $ \Lambda =+1,\, 0 $.
Thus one obtains a set of two coupled integral equations in two variables,
the magnitude of the relative momentum vector, i.e. $p$, and the angle
between ${\bf p}$ and the arbitrarily chosen $z$-axis, i.e. $\theta$.

\section{Preparation of the Chiral Potential in 3D representation} \label{sec:chiral potential}

The general form of the $NN$ potential by considering the rotation, parity, and time reversal invariance can be written as a linear combination of six $\Omega_i$ operators, which are consistent with the helicity basis representation \cite{Fachruddin-PRC69}:
\begin{eqnarray}
 \langle{\bf p}'|V|{\bf p}\rangle\equiv V({\bf p}',{\bf p})=\sum^{6}_{i=1}v_{i}(p',p,\gamma)\, \sum_{j}A_{ij}\Omega_{j},
\end{eqnarray}
where $v_{i}(p',p,\gamma)$ are scalar functions which depend on the magnitudes of
${\bf p}$ and ${\bf p}'$ and also the angle between them,
$\gamma=\hat{p}.\hat{p}'$, and $A$ is a $6\times6$ matrix. The $\Omega_i$ operators are:
$\Omega_{1} = 1, \Omega_{2} = {\bf S}^{2}, \Omega_{3} = {\bf S}\cdot \hat{p}'{\bf S}\cdot \hat{p}',
\Omega_{4} = {\bf S}\cdot \hat{p}'{\bf S}\cdot \hat{p}, \Omega_{5} = ({\bf S}\cdot \hat{p}')^{2}({\bf S}\cdot \hat{p})^{2},
\Omega_{6} = {\bf S}\cdot \hat{p}{\bf S}\cdot \hat{p}.$ By this representation the spin-dependent parts of the matrix elements of the potential can be easily evaluated in the helicity basis states.
We intend to use the chiral $NN$ potential up to N$^3$LO
of chiral expansion which consists of "one- and two-pion exchanges (1PE, 2PE) and a string of contact interactions with an increasing number of derivatives (zero, two, four) that parameterize the shorter ranged components of the nuclear force" \cite{Epelbaum-NPA747}.
In order to use the chiral potential in 3D formalism we need first to rewrite this potential in an appropriate operator form which is consistent with helicity representation. To this aim we should overcome the following two possible issues:
  \begin{itemize}
   \item the calculation of original low energy coefficients (LEC's) for incorporating the contact terms
   \item the representation of the spin dependent parts in term of $\Omega_i$ operators
  \end{itemize}

  As indicated in reference \cite{Epelbaum-NPA747} the chiral potential at N$^3$LO consists of contact terms which contain 24 original LEC's; $C_S$, $C_T$, $C_1$, ..., $C_7$ and $D_1$, ..., $D_{15}$.
  In order to calculate any observable with chiral potential in the PW approach it is sufficient to project only the contact interactions in the fourteen PW channels up to $J=3$ and it is not necessary to consider the higher channels, i.e. $J=4, 5, 6$, etc.
Once the 24 spectroscopic LEC's have been determined by fitting to the phase shifts of the Nijmegen potential, the original ones can be obtained uniquely.
This is a serious problem to apply the contact terms of the chiral potential in 3D approach,
since in this approach we consider all of the PW channels
simultaneously. To overcome this problem we have used the connection between the 3D and PW representations of matrix elements of the $NN$ potential \cite{Fachruddin-PhD} to sum over these fourteen channels and to obtain the matrix elements of the potential in momentum helicity basis.

In order to make the chiral potential compatible with helicity representation, we highlight the spin dependent parts of the potential as follows:
\begin{eqnarray}
&& \fet \sigma_1 \cdot \fet \sigma_2, \quad \fet \sigma_1 \cdot {\bf q}\, \fet \sigma_2 \cdot {\bf q}, \quad \fet \sigma_1 \cdot {\bf k}\, \fet \sigma_2 \cdot {\bf k}, \nonumber \\
&& i(\fet \sigma_1 + \fet \sigma_2)\cdot({\bf q}\times {\bf k}), \quad \fet \sigma_1 \cdot ({\bf q}\times {\bf k})\, \fet \sigma_2\cdot({\bf q}\times {\bf k}).
\end{eqnarray}
These parts can be simply written in term of $\Omega_i$ operators as:
\begin{eqnarray}
\fet \sigma_1 \cdot \fet \sigma_2 = 2\Omega_2-3\Omega_1,
\end{eqnarray}
\begin{eqnarray}
 \fet \sigma_1 \cdot {\bf q}\, \fet \sigma_2\cdot
{\bf q} = -b\Omega_1+\frac{p'p\,a^2}{\gamma}\Omega_2+\frac{2p'(p'\gamma-p)}{\gamma }\Omega_3
\nonumber \\ - 2p'p\Omega_4+\frac{2p'p}{\gamma}\Omega_5+\frac{2p(p\gamma-p')}{\gamma
}\Omega_6,
\end{eqnarray}
\begin{eqnarray}
\fet \sigma_1 \cdot {\bf k} \, \fet \sigma_2\cdot
{\bf k} = -\frac{c}{4}\Omega_1-\frac{p'p\,a^2}{4\gamma
}\Omega_2+\frac{p'(p'\gamma+p)}{2\gamma }\Omega_3
\nonumber \\
+\frac{p'p}{2}\Omega_4-\frac{p'p}{2\gamma}\Omega_5+\frac{p(p\gamma+p')}{2\gamma
}\Omega_6,
\end{eqnarray}
\begin{eqnarray}
i(\fet \sigma_1+ \fet \sigma_2)\cdot({\bf q}\times
{\bf k}) \equiv i(\fet \sigma_1+ \fet \sigma_2)\cdot({\bf p}'\times {\bf p})
\nonumber \\ =  \frac{p'p\,a^2}{\gamma}\Omega_2-\frac{2p'p}{\gamma
}\Omega_3+2p'p\Omega_4+\frac{2p'p}{\gamma
}\Omega_5+\frac{2p'p}{\gamma }\Omega_6,
\end{eqnarray}
\begin{eqnarray}
\fet \sigma_1 \cdot ({\bf q}\times
{\bf k})\, \fet \sigma_2\cdot({\bf q}\times
{\bf k}) \equiv \fet \sigma_1 \cdot ({\bf p}'\times
{\bf p})\, \fet \sigma_2\cdot({\bf p}'\times {\bf p})
\nonumber \\
= -p'^2p^2a^2\Omega_1+p'^2p^2a^2\Omega_2+2p'^2p^2\gamma\Omega_4-2p'^2p^2\Omega_5,
\end{eqnarray}
where $ \gamma = \hat{p}'\cdot \hat{p}, \, a = \sqrt{1-\gamma^2}, \, b = p'^2+p^2-2p'p\gamma=q^2, \, c = p'^2+p^2+2p'p\gamma=4k^2$.

\section{Numerical results}\label{sec:results}
In this section we present the obtained numerical results for np differential cross section and deuteron binding energy with chiral potential up to N$^3$LO in the 3D approach. In order to demonstrate the effectiveness of 3D formalism we have compared
our numerical results with the corresponding PW results as well as the experimental data. The low energy coefficients in the chiral potential are determined for given cut-off parameters $\Lambda_1$ and $\Lambda_2$ by fitting to $NN$
data, where the cut-off $\Lambda_1$ regulates the high-momentum components of the interacting nucleons and the cut-off $\Lambda_2$ which appears in the spectral function regularization excludes the
high-momentum components of the two-pion exchange. We use in our
calculations different combinations of ($\Lambda_1, \, \Lambda_2$) for N$^3$LO as: (450,500), (600,600), (550,600), (450,700), (600,700) in units of MeV/c.
\begin{table}[hbt]
\caption {Deuteron binding energy calculated for the chiral potential at N$^3$LO in 3D approach for three different cut-off sets in comparison with the PW and experimental results.}
\begin{tabular}{ccccccccccccccccccccc}
\hline \hline
$(\Lambda_1,\Lambda_2)$ &&& 3D [MeV] &&& PW \cite{Young} [MeV] \\
\hline
(450,500) &&& -2.216 &&& -2.215 \\
(450,700) &&& -2.219 &&& -2.218 \\
(600,700) &&& -2.222 &&& -2.220 \\
\hline
PW \cite{Epelbaum-NPA747} &&&&& (-2.216)-(-2.223) \\
EXP &&&&& -2.224575(9) \\
\hline \hline
\end{tabular}
\label{table.deuteron}
\end{table}

\begin{figure*}[hbt]
\leftskip=-8cm
\includegraphics[width=40cm]{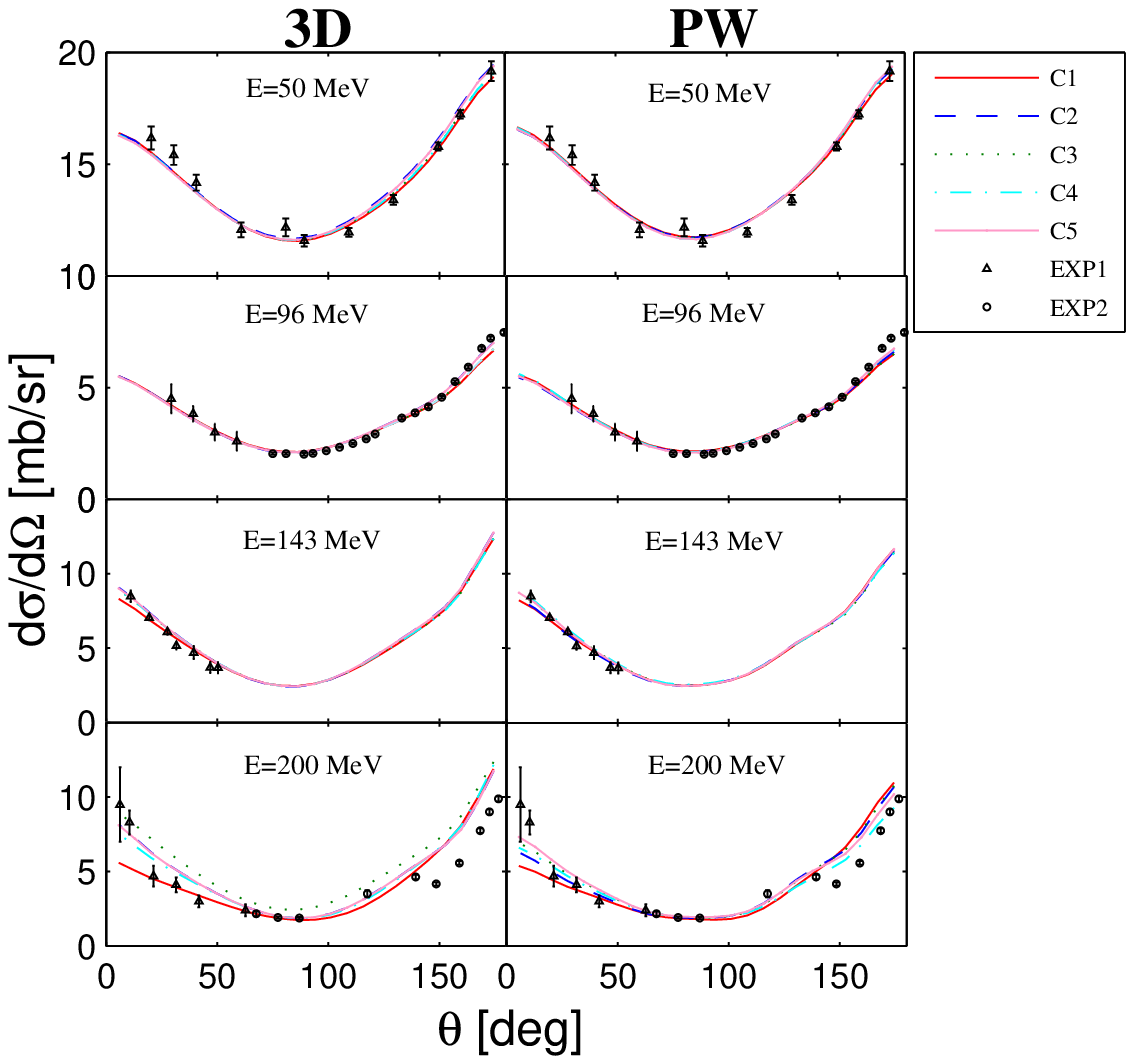}
\caption{The np differential cross section for four different energies. The left figures are 3D and the right ones are PW results.
In each figure the 3D or PW numerical results are obtained
for five cut-off sets, i.e. C1:(450,500), C2:(600,600), C3:(550,600), C4:(450,700), C5:(600,700) MeV.
The PW results have been obtained up to $J=6$ and they have been taken from ref. \cite{Epelbaum-NPA747}.
The experimental data have been taken from ref. \cite{Montgomery-PRC16} for $E_{lab}=50$ MeV (EXP1), from
refs. \cite{Griffith-PPSL71, Rahm-PRC63} for $E_{lab}=96$ MeV (EXP1 and EXP2 respectively), from
ref. \cite{Bersbach-PRD13} for $E_{lab}=142.8$ MeV (EXP1) and from refs. \cite{Hurster-PLB90, Kazarinov-SP16} for $E_{lab}=200$ MeV (EXP1 and EXP2 respectively).}
\label{Figure:np}
\end{figure*}

The numerical results for np differential cross section in four different energies of the
projectile in the laboratory system and for different cut-off sets have been shown in figure (\ref{Figure:np}). In the first row of this figure we have presented a comparison between 3D and PW
results for $E_{lab}=50$ MeV. The PW results have been taken from
\cite{Epelbaum-NPA747} where the calculation is up to $J_{max}=6$.
Both 3D and PW results are in good agreement in forward and backward angles.
Also the comparison of both approaches with experimental
data in backward angles shows a very close agreement. In the next rows the same comparison has been shown for
$E_{lab}=96, 143$ and $200$ MeV, and as we see both approaches match together and also to the experimental data perfectly.
In table \ref{table.deuteron} we have presented our numerical results for deuteron binding energy in comparison with PW and experimental data. Our numerical results for three cut-off sets with the values $-2.216, \, -2.219$ and $-2.222$ MeV are in good agreement with the very recent corresponding PW results \cite{Young}, and also with Epelbaum \emph{et al.} PW achievements \cite{Epelbaum-NPA747}. The agreement between the 3D and PW results as well as the experimental data is quiet satisfactory.

Although we have studied the $2N$ systems we conclude that the 3D approach is promising to be simpler for more complex few-body systems by providing a strictly finite number of coupled three-dimensional integral equations to be solved.
The number of the integral equations in the 3D approach is consistent and do not depend on the
energy of the system. This subject is more important when we
consider $3N$ and $4N$ scattering problems in which the number of the equations in
the higher energies makes the problems more complex. The recently developed $3N$ bound state in 3D approach \cite{Bayegan-PRC77} can be used to calculate the $^3$H and $^3$He binding energies by using the chiral potential. The $3N$ scattering, Nd capture and $3N$ photodisintegration calculations with this new form of the chiral potential are interesting goals that we are pursuing. Also the incorporation of $3N$ chiral forces in novel 3D approach calculations is one of other interesting problems that can be done.

\section*{Acknowledgments}
We would like to highly appreciate E. Epelbaum for providing us the chiral $NN$ partial wave code. We also convey special thank to W. Gl\"{o}ckle, Ch. Elster and I. Fachruddin for application of their helicity formalism. This work was supported by the research council of the University of Tehran.

\end{document}